\documentclass[aps,prb,groupedaddress,nofootinbib]{revtex4}
\usepackage{graphicx}
\usepackage{hyperref}
\usepackage[english]{babel}
\usepackage{makeidx}
\usepackage{bm}
\usepackage{color}

\newcommand{\bfr}{{\bf r}}

\newcommand{\Ekin}{T}

\newcommand{\rsup}{r_{>}}
\newcommand{\rinf}{r_{<}}
\newcommand{\absv}[1]{\left|\displaystyle #1\right|}

\newcommand*{\bra}[1]{\ensuremath{\left\langle #1 \right|}}
\newcommand*{\ket}[1]{\ensuremath{\left| #1 \right\rangle}}
\newcommand*{\norm}[1]{\ensuremath{\left| #1 \right|}}
\newcommand{\braa}{\left[}
\newcommand{\kett}{\right]}
\newcommand{\balpha}{\bm{\alpha}}
\newcommand{\bmu}{\bm{\mu}}
\newcommand{\bdelta}{\bm{\delta}}
\newcommand{\bnabla}{\bm{\nabla}}
\newcommand{\bPi}{\bm{\Pi}}

\begin{document}

\title{Including many-body effects in models for ionic liquids}

\author{Mathieu Salanne$^{1}$}
\author{Benjamin Rotenberg$^{1}$}
\author{Sandro Jahn$^2$}
\author{Rodolphe Vuilleumier$^3$}
\author{Christian Simon$^1$}
\author{Paul A. Madden$^4$}

\affiliation{$^1$ UPMC Univ Paris 06, CNRS, ESPCI, UMR 7612, PECSA, F-75005 Paris, France}
\affiliation{$^2$ GeoForschungsZentrum Potsdam, Section 4.1, Telegrafenberg, 14473 Potsdam, Germany}
\affiliation{$^3$ ENS, CNRS, UMR 8640, PASTEUR, F-75005, Paris, France}
\affiliation{$^4$ Department of Materials, University of Oxford, Parks Road, Oxford OX1 3PH, United Kingdom}



\date{\today}

\begin{abstract}
Realistic modeling of ionic systems  necessitates taking explicitly account of many-body effects. In molecular dynamics simulations, it is possible to introduce explicitly these effects through the use of additional degrees of freedom. Here we present two models: The first one only includes dipole polarization effect, while the second also accounts for quadrupole polarization as well as the effects of compression and deformation of an ion by its immediate coordination environment. All the parameters involved in these models are extracted from first-principles density functional theory calculations. This step is routinely done through an extended force-matching procedure, which has proven to be very succesfull for molten oxides and molten fluorides. Recent developments based on the use of localized orbitals can be used to complement the force-matching procedure by allowing for the direct calculations of several parameters such as the individual polarizabilities. 
\end{abstract}
\maketitle

\section{Introduction}

Ionic liquids  are a class of coulombic liquids~\cite{hansen-livre} which encompasses several types of compounds. They are often separated into just two categories, inorganic molten salts and room-temperature ionic liquids, although this may be too restrictive. For example, inorganic molten salts contain various families based on monoatomic anions such as molten halides or molten oxides, but also other ones in which molecular anions are involved such as carbonates and nitrates. In the case of room-temperature ionic liquids, organic cations are involved, which opens the way for an almost infinite number of potential electrolytes exhibiting a wide range of physical and chemical properties.~\cite{hallett2011a}

Ionic liquids play an important role in many fields of chemistry and physics. Apart from the intrinsic interest for such systems, they have been studied by computer simulations in many contexts. For example, molten oxides are the main components of magmatic melts~\cite{jahn2007b,guillot2007a,guillot2007b,vuilleumier2009a}, molten fluorides and chlorides are investigated for their importance in many metallurgical processes~\cite{chen2000a,elghallali2009a,groultasap} and for their potential use in the nuclear industry~\cite{salanne2006b,salanne2008a}, and room temperature ionic liquids for their use in electrochemical storage applications~\cite{siqueira2007a,borodin2009a,vatamanu2011a} or their solvation properties~\cite{maginn2007a,zhao2009a,kohagen2011a}.

 In order to obtain the thermodynamic and transport properties of interest for all these applications, it is necessary to simulate large enough samples for a sufficiently long time, which is out of reach of {\it ab initio} molecular dynamics methods. This is done instead via classical simulations, in which interactions are described by an analytical force field derived from a model of the interactions between closed-shell species. Such a model~\cite{stone2008a} must account not only for the classical electrostatic interaction, but also for three interactions arising from the quantum nature of electrons. The exchange-repulsion, or van der Waals (VdW) repulsion is a consequence of the Pauli principle, while the dispersion (VdW attraction) arises from correlated fluctuations of the electrons. Lastly, the induction term reflects the distortion of the electron density in response to electric fields, which are dominated by polarization effects but may also partially represent {\it incipient} charge transfer associated with bond formation between closed-shell species.~\cite{ribeiro2000a} 

In this paper we  review some methodological advances which have been proposed in order to include all these effects, and particularly the many-body polarization ones, in accurate models of ionic liquids. The critical physical factor to appreciate is that the electronic densities of anions, in particular, are strongly affected by their interactions with their environment. For example, the oxide ion, which as an isolated ion is unstable by 8~eV with respect to autoionization \cite{harding1995a}, only exists in the condensed phase because of the confining potential exerted on its electrons by its neighbours \cite{fowler1984a,fowler1985a}. The confining potential compresses and stabilizes the oxide ion and allows us to treat its interactions in the condensed phase as those of a closed-shell species, like an inert-gas atom. However a model which is {\it transferable} from one material to another, or even between different phases of the same material, must recognise that the confining potential, and hence the intrinsic properties of the oxide ion will change between them. Furthermore, in a dynamical context,  fluctuations in the environmental potential due to the thermal motion of the neighbours mean that the ion's electron density is also fluctuating and this effect must be included in the interaction model. The effect is less severe for halide ions, where the isolated ion is stable, but where the importance of the environmental effect is indicated by the observation that the in-crystal polarizability of the fluoride ion in crystalline LiF is 6.2~a.u. compared  16~a.u. for the gas-phase ion.~\cite{fowler1984a,fowler1985a} One would expect that for the large molecular anions involved in room temperature systems, the effect would be weaker still.    Our considerations will be directed at the inorganic molten salt systems for which the environmental potential effects are pronounced, but with the compensating simplification that the complexities of molecular shape are avoided. However, we note that similar methods have recently been applied to develop polarizable models for room temperature systems.~\cite{salanneinpress}

In the first part of the present paper, we provide the expression for the interaction potential in the framework of two models with different complexity: the aspherical ion model, necessary for a transferable description of oxides, and the simpler polarizable ion model. We then show how all the parameters involved in these models can be determined from first-principles density functional theory calculations through an extended force-matching procedure. In a third part, we show recent developments based on the use of localized orbitals, which can complement the force-matching procedure by allowing for the direct calculations of several parameters such as the individual polarizabilities.

\section{Interaction potentials}
The functional forms introduced below have been proposed previously on the basis of separate examinations of each contribution to the ionic interaction energy in a series of well-directed electronic structure calculations on the condensed-phase.~\cite{wilson1996c,rowley1998a,jemmer1999a} The potential is best described as the sum of four different components: charge-charge, dispersion, overlap repulsion and polarization.
\begin{equation}
V^{\rm total}=V^{\rm charge}+V^{\rm dispersion}+V^{\rm repulsion}+V^{\rm polarization}
\end{equation}

Two of these terms will attract more our attention, the repulsion and polarization terms.
These contributions to the overall energy depend on ionic
properties like the ionic radius, for the repulsion, or the ionic
multipoles, for the polarization. As the electronic density of the
ions adapt to the environment that act on it through a confining
potential or local electric fields, these ionic quantities change
in a way so as to minimize the overall energy of the system. The
response to the environment of these additional degrees of freedom such
as ionic radius and multipoles gives rise to many-body
effects in the potential energy of the system. 

The first term corresponds to the electrostatic interaction between two formal charges,
\begin{equation}
V^{\rm charge}=\sum_{i<j}\frac{q^iq^j}{r^{ij}}
\end{equation}
\noindent where $q^i$ is the charge of ion $i$. The use of formal charges underpins the description of the other interactions as characteristic of a closed-shell system~\cite{stone-book} and the expectation that the interaction parameters should be transferable. The dispersion component includes dipole-dipole and dipole-quadrupole terms

\begin{equation}
V^{\rm dispersion}=-\sum_{i<j}\left(f_6^{ij}(r^{ij})\frac{C_6^{ij}}{(r^{ij})^6}+f_8^{ij}(r^{ij})\frac{C_8^{ij}}{(r^{ij})^8} \right)
\label{eq:dispersion}
\end{equation}

\noindent where $C_6^{ij}$ ($C_8^{ij}$) is the dipole-dipole (dipole-quadrupole) dispersion coefficient, and $f_n^{ij}$ are Tang-Toennies dispersion damping functions,~\cite{tang1984a} describing the short-range penetration correction to the asymptotic multipole expansion of dispersion,~\cite{stone-book} which take the following form:

\begin{equation}
f_n^{ij}(r^{ij})=1-{\rm e}^{-b_n^{ij}r^{ij}}\sum_{k=0}^n\frac{(b_n^{ij}r^{ij})^k}{k!}
\label{eq:tangtoennies1}
\end{equation}

\noindent where the $b_n^{ij}$ parameter sets the range of the damping effect. The two other terms  which include many-body effects may differ depending on the systems of interest, as the magnitude of the response to the environment varies from one species to another. In the most complex functional form, the so-called ``aspherical ion model'' (AIM), which has been introduced to model oxide materials,~\cite{aguado2003b,aguado2004a,jahn2007b} the overlap repulsion component is given by

\begin{eqnarray}
V^{\rm repulsion}_{\rm AIM}&=&\sum_{i\in {\rm O},j\in {\rm M}}\left( A^{ij}{\rm e}^{-a^{ij}\rho^{ij}}+B^{ij}{\rm e}^{-b^{ij}\rho^{ij}}+C^{ij}{\rm e}^{-c^{ij}r^{ij}}\right) \\
& &+ \sum_{i,j\in{\rm O}, i<j}A^{ij}{\rm e}^{-a^{ij}r^{ij}}+\sum_{i,j\in{\rm M}, i<j}A^{ij}{\rm e}^{-a^{ij}r^{ij}}  \nonumber \\
& &+\sum_{i\in{\rm O}} \left[D^i({\rm e}^{\beta^i\delta\sigma^i}+{\rm e}^{-\beta^i\delta\sigma^i}) +({\rm e}^{\zeta^i\mid\nu^i\mid^2}-1)+({\rm e}^{\eta^i\mid\kappa^i\mid^2}-1)\right] \nonumber
\label{eq:repulsion1}
\end{eqnarray}

\noindent where
\begin{eqnarray}
\rho^{ij}=r^{ij}-\delta\sigma^i-S_\alpha^{(1)}\nu_\alpha^i-S_{\alpha\beta}^{(2)}\kappa_{\alpha\beta}^i,
\end{eqnarray}
\noindent and summation of repeated indices is implied. Sets $\rm O$ and $\rm M$ are the sets of oxide anions and cations respectively, so that the first three summations represent cation-anion, anion-anion and cation-cation short-range repulsions, respectively. In the cation-anion term, $\delta\sigma^i$ is a variable which characterizes the deviation of the radius of the oxide anion $i$ from its default value, {\it i.e.} its ``breathing''. $\{\nu_\alpha^i\}$ is a set of three variables, for each anion, describing the Cartesian components of a dipolar distortion of the ion shape. Similarly, $\{\kappa_{\alpha\beta}^i\}$ is a set of five independent variables describing the corresponding quadrupolar shape distortions. $\mid\kappa\mid^2=\kappa_{xx}^2+\kappa_{yy}^2+\kappa_{zz}^2+2(\kappa_{xy}^2+\kappa_{xz}^2+\kappa_{yz}^2)$, and the two interaction tensors $S_\alpha^{(1)}=r_\alpha^{ij}/r^{ij}$ and $S_{\alpha\beta}^{(2)}=3r_\alpha^{ij}r_\beta^{ij}/(r^{ij})^2-\delta_{\alpha\beta}$ are also used in equation \ref{eq:repulsion1}. The last summations include the self-energy terms, that is the energy cost of deforming the charge density of an ion. $\beta$, $\zeta$ and $\eta$ are effective force constants determining how difficult it is for a particular ion to be deformed in a spherical, dipolar or quadrupolar way.

The magnitude and orientation of the spherical breathing, dipolar and quadrupolar deformations will be determined by minimization of this energy, with respect to these variables: they will therefore depend on the instantaneous positions of neighboring ions and consequently change at each timestep in a molecular dynamics run.

It was assumed throughout that these shape deformations do not affect significantly the anion-anion and cation-cation repulsions, which have been represented by simple exponentials as if the ions were effectively spherical in their interactions with other ions of the same type. We have also assumed that short range cation distortion effects can be neglected in the case of oxide materials.

The polarization part of the AIM potential includes both dipolar and quadrupolar contributions

\begin{eqnarray}
V^{\rm polarization}_{\rm AIM}&=&\sum_{i,j}\left[ \left(q^i\mu^j_{\alpha}g_D^{ij}(r^{ij})-\mu_\alpha^iq^jg_D^{ji}(r^{ij})\right)T_\alpha^{(1)}  \right. \\
 & & \left. + \left(\frac{q^i\theta^j_{\alpha\beta}}{3}g_Q^{ij}(r^{ij})+\frac{\theta^i_{\alpha\beta}q^j}{3}g_Q^{ji}(r^{ij})-\mu_\alpha^i\mu_\beta^j\right)T_{\alpha\beta}^{(2)} \right. \\
 & & \left. + \left(\frac{\mu_\alpha^i\theta^j_{\beta\gamma}}{3}+\frac{\theta^i_{\alpha\beta}\mu_\gamma^j}{3}\right)T_{\alpha\beta\gamma}^{(3)}+\frac{\theta^i_{\alpha\beta}\theta_{\gamma\delta}^j}{9}T_{\alpha\beta\gamma\delta}^{(4)}\right] \nonumber \\
 & & +\sum_i \left(k_1^i\mid\bmu^i\mid^2+k_2^i\mu_\alpha^i\theta_{\alpha\beta}^i\mu_\beta^i+k_3^i\theta^i_{\alpha\beta}\theta^i_{\alpha\beta} +k_4^i \mid \bmu^i \cdot \bmu^i\mid^2 \right)
\end{eqnarray}
where $k_1^i=\frac{1}{2\alpha^i}$, $k_2^i=\frac{B^i}{4(\alpha^i)^2C^i}$, $k_3^i=\frac{1}{6C^i}$ and $k_4^i=\frac{-(B^i)^2}{16(\alpha^i)^4C^i}$. $\alpha^i$, $B^i$ and $C^i$ are the dipole, dipole-dipole-quadrupole and quadrupole polarizabilities of ion $i$. $T_{\alpha\beta\gamma...}=\nabla_\alpha \nabla_\beta \nabla_\gamma\cdot\cdot\cdot(r^{ij})^{-1}$ are the multipole interaction tensors,~\cite{stone-book} with the superindex indicating the order of the operator. They are computed using the Ewald summation technique.~\cite{aguado2003a,laino2008a} The instantaneous values of the dipole and quadrupole moments are, again, obtained by minimization of this expression. The charge-dipole and charge-quadrupole cation-anion asymptotic terms are also corrected for penetration effects~\cite{wang2010b} at short-range by using Tang-Toennies damping functions~\cite{tang1984a}

\begin{eqnarray}
g_D^{ij}(r^{ij})=1-c_D^{ij}{\rm e}^{-b_D^{ij}r^{ij}}\sum_{k=0}^n\frac{(b_D^{ij}r^{ij})^k}{k!} \\
g_Q^{ij}(r^{ij})=1-c_Q^{ij}{\rm e}^{-b_Q^{ij}r^{ij}}\sum_{k=0}^n\frac{(b_Q^{ij}r^{ij})^k}{k!}
\end{eqnarray}

\noindent with $D$ and $Q$ standing for the dipolar and quadrupolar parts. Note that the $b_D$ ($b_Q$) parameters, which determine the distance at which the overlap of the charge densities begins to affect the induced multipoles, are the same for both anions and cations, while the $c_D$ ($c_Q$) parameters, which are set to unity in the case of the dispersion interaction (equation \ref{eq:tangtoennies1}), measure the strength of the ion response to this effect and therefore depend on the identity of the ion. The necessity of larger-than-unity values was indicated by {\it ab initio} calculations of the induced multipoles in distorted crystals.~\cite{jemmer1999a} The short range induction corrections are usually neglected in both anion-anion and cation-cation interactions.

The AIM potential can be seen to contain several (seventeen per oxide anion) additional degrees of freedom (induced dipoles and quadrupoles, and ion shape deformations), which describe the state of the electron charge density of the ions. When calculating the forces on the ions in a molecular dynamics simulation, these electronic degrees of freedom should have their adiabatic ``Born-Oppenheimer'' values, which minimize the total potential energy, for every atomic configuration. The values taken by these extra degrees of freedom
is then a complicated functional of the atomic configurations so
that, even if each term in the total energy is written as a sum of
individual or pair components, the total system energy includes
many-body effects. We search for the ground state configuration of the electronic degrees of freedom at each time step, using a conjugate gradients routine, i.e.
\begin{eqnarray}
\left(\frac{\partial V^{\rm repulsion}_{\rm AIM}}{\partial \xi^j}\right)_{\{\xi^M\}}=0, \hspace{0.5cm}{\rm where } \hspace{0.5cm} \{\xi^M\}=\{\delta\sigma^N,\nu_\alpha^N,\kappa_{\alpha\beta}^N\}  \\
\left(\frac{\partial V^{\rm polarization}_{\rm AIM}}{\partial \xi^j}\right)_{\{\xi^M\}}=0, \hspace{0.5cm} {\rm where } \hspace{0.5cm} \{\xi^M\}=\{\mu_\alpha^N, \theta_{\alpha\beta}^N\}  \\
\end{eqnarray}

\noindent  The convergence criteria typically require the energy at successive steps to differ only in the 8th significant figure for polarization and in the 10th for repulsion. The dynamics is thus similar to the so-called Born-Oppenheimer {\it ab initio} molecular dynamics.  Since the ions remained fixed during the conjugate gradient process many of the terms required to evaluate the energy do not need to be recalculated at each minimization step.

 In the case of halide systems, we observed that a simpler functional form was sufficient to describe accurately the interactions: In that case, we have used the ``polarizable ion model'' (PIM), where the repulsion term does not include any ion shape deformation term anymore, which corresponds to the simple exponential form:

\begin{eqnarray}
V^{\rm repulsion}_{\rm PIM}&=&\sum_{i<j} A^{ij}{\rm e}^{-a^{ij}r^{ij}}
\end{eqnarray}

\noindent In addition, good representations of the halides are found with the polarization term only including the terms up to the dipoles,
\begin{eqnarray}
V^{\rm polarization}_{\rm PIM}&=&\sum_{i,j}\left[ \left(q^i\mu^j_{\alpha}g_D^{ij}(r^{ij})-\mu_\alpha^iq^jg_D^{ji}(r^{ij})\right)T_\alpha^{(1)}-\mu_\alpha^i\mu_\beta^jT_{\alpha\beta}^{(2)}\right] \\
& & +\sum_i \left(\frac{1}{2\alpha^i}\mid \bmu^i \mid^2 \right) \nonumber
\end{eqnarray}

\noindent With these simplifications, the potential now includes only three additional degrees of freedom per ion, associated with the induced dipoles, which are calculated in a single minimization procedure:
\begin{eqnarray}
\left(\frac{\partial V^{\rm polarization}_{\rm PIM}}{\partial \mu_\alpha^i}\right)_{\{\mu_\alpha^N\}}=0  \\
\end{eqnarray}

 Such interaction potentials have been introduced by our group in the case of molten chlorides~\cite{wilson1993a,wilson1994a} and fluorides.~\cite{foy2006a,heaton2006a,salanne2009a} They were also used by Trullas and co-worker to study a series of silver and copper halides,~\cite{bitrian2006a,bitrian2007a,bitrian2008a,alcaraz2011a,bitrian2011a} and the PIM model now is implemented in the CP2K code.~\cite{cp2k}  Going back to the case of oxides, when attention is restricted to a single phase or where similar materials are being compared it is often sufficient to neglect the full complexity of the AIM model, and we have also successfully used some PIM-type potentials in the case of amorphous GeO$_2$~\cite{marrocchelli2009a,marrocchelli2010a} and of doped zirconia crystals.~\cite{norberg2009a,marrocchelli2009b,norberg2011a,marrocchelli2011a} The potential developed for SiO$_2$ by Tangney and Scandolo~\cite{tangney2002a} on the basis of similar force-matching methods to those we describe below is also worth noting. Their model, which performs very nicely in reproducing the properties of various phases of SiO$_2$,~\cite{liang2006a,giacomazzi2011a} differs from the PIM one due to the use of partial charges for the ions, which hinders its transferability to other silicate compounds.

\section{Obtaining the parameters from force-matching and multipole-matching}

In their pioneering work, Tosi and Fumi developed a series of force field parameters for alkali halides in the framework of the Born model.~\cite{fumi1964a,tosi1964a} These parameters were chosen in order to reproduce solid state data. Their potential have then been used in many simulation works involving alkali halides in the molten state.~\cite{lantelme1974a,lantelme1982a,anwar2003a} This type of empirical approach has long been the standard procedure for parameterizing interaction potentials. A new source of data for fitting these potentials was made possible by the development of {\it ab initio} techniques. For example, the most commonly used potential for SiO$_2$ is based on Hartree-Fock calculations of a single H$_4$SiO$_4$ cluster and on several crystal properties.~\cite{vanbeest1990a} Oeffner and Elliott used a similar approach to develop an interaction potential for GeO$_2$.~\cite{oeffner1998a}

The introduction of the Car-Parrinello method~\cite{car1985a} opened the way to a new class of simulations, namely the {\it ab initio} molecular dynamics (AIMD).~\cite{editorialcarparrinello} In AIMD simulations, may it be of the Car-Parrinello or of the Born-Oppenheimer type, the introduction of an analytical interaction potential is avoided through the use of density functional theory (DFT) calculations. This has an appealing ``hands free'' aspect, but unfortunately these methods cannot yet access the time scales necessary to determine many material properties, in particular in ionic liquids transport properties are of primary importance. Nevertheless, AIMD simulations are now considered to be very accurate, and they can be used as a reference for classical MD simulations. The idea has therefore emerged to develop interaction potentials yielding a trajectory that {\it mimics} the one which would be obtained by AIMD. Since molecular dynamics is based on an iterative integration of Newton's equation of motion, this can be done if the forces generated by the interaction potential are the same as the DFT-calculated ones. This approach,~\cite{laio2000a} which is now termed force-matching or force-fitting, has been used (and extended) for a decade to develop accurate force fields for ionic liquids in the framework of the PIM and of the AIM. We will now focus on practical details of these parameterizations.

DFT calculations on periodic systems are based on the Kohn-Sham (KS) method
to determine the ground state electronic wavefunction $\{\phi^0\}$ of
a given condensed phase configuration containing $N$ ions. This wavefunction is obtained by minimization of the Kohn-Sham energy $E^{KS}$. The force acting on each atom $i$, with position ${\bf r}^i$, is then extracted from:
\begin{equation}
F^i_{\alpha, {\rm DFT}}=\frac{\partial E^{KS}[\{\phi^0\}]}{\partial r_\alpha^i}
\end{equation}

The corresponding classical forces for the same condensed phase configuration are easily obtained for a given set of parameters from:
\begin{equation}
F^i_{\alpha, {\rm classical}}=-\frac{\partial V^{\rm total}}{\partial r_\alpha^i}
\end{equation}
\noindent where the interaction potential could either derive from the PIM or the AIM, and the additional degrees of freedom minimize the total energy for this configuration. The idea behind force-fitting consists of finding the set of parameters which will minimize the error made in the classical calculation with respect to the DFT one. If this error is minimal, the interaction potential can be considered to be of {\it ab initio} accuracy. An important concern when dealing with ionic systems is that it is often required to perform studies with varying conditions of temperature, pressure and compositions. A prerequisite of the interaction potentials therefore is to be {\it transferable} from one thermodynamic point to another, which can be achieved by determining the interaction potential parameters on a series of representative condensed phase configurations instead of one. In practice, this is done by minimizing the following expression:
\begin{equation}
\chi_F^2=\frac{1}{N_c}\sum_{j=1}^{N_c}\frac{1}{N_j}\sum_{i=1}^{N_j}\frac{\mid{\bf F}^i_{\rm DFT}-{\bf F}^i_{\rm classical}\mid^2}{\mid{\bf F}^i_{\rm DFT}\mid^2}
\label{eq:chi2forces}
\end{equation}

\noindent where $N_c$ is the number of configurations and $N_j$ is the number of ions in configuration $j$. Another quantity that can also be determined rather straightforwardly from both the {\it ab initio} and the classical calculation is the stress tensor\footnote{In the case of a classical calculation involving additional degrees of freedom, the correct expression for the stress tensor (as well as for the heat current) is obtained by exploiting the Hellmann-Feynman theorem and ignoring any explicit derivatives of the additional degrees of freedom with respect to the ion positions.}. It is therefore possible to extend the fitting dataset by minimizing
\begin{equation}
\chi_S^2=\frac{1}{N_c}\sum_{j=1}^{N_c}\frac{\sum_{\alpha\beta}\mid\Pi^i_{\alpha\beta,{\rm DFT}}-\Pi^i_{\alpha\beta,{\rm classical}}\mid^2}{\sum_{\alpha\beta}\mid\Pi^i_{\alpha\beta,{\rm DFT}}\mid^2}
\label{eq:chi2stress}
\end{equation}

In that case relative weights then have to be assigned to $\chi_F^2$ and $\chi_S^2$ in the minimization procedure. Of course, all the parameters of the interaction potential are involved in the calculation of the force and of the stress tensor, so that they can all be fitted at this stage. There is nevertheless a risk of interplay between the various terms of the potential, so that one term will do the work which should be attributed to the other. This would contribute to a loss of transferability of the potential.

Consequently, we generalize the force-fitting to require that the induced multipoles on our ions in the model reproduce the {\it ab initio} ones and determine the parameters in $V^{\rm polarization}$ from this requirement. In DFT calculations it is possible to calculate the multipole moments of the individual ions thanks to the maximally localized Wannier function (MLWF)
formalism.~\cite{marzari1997a} The MLWFs provide a picture of the electron distribution around
atoms which is easily interpreted from a chemical
point of view. They are determined by unitary transformations of the KS
eigenvectors
\begin{equation}
\ket{\phi^w_n}=\sum_{m=1}^N U_{nm}\ket{\phi_m^0}
\end{equation}
where the sum runs over all the occupied KS states
$\{\phi_n\}_{n\in[1,N]}$, and the unitary matrix ${\bm U}$ is determined by
iterative minimization of the Wannier function spread $\Omega$, which is
defined as
\begin{equation}
  \Omega = -\frac{1}{(2\pi)^2}{\sum_{n=1}^N \log \norm{{\bm{s}}_n}^2}\,; \quad
  s_{n,\alpha} = \bra{\phi_n^w}e^{-i \frac{2\pi}{L} r_\alpha}\ket{\phi_n^w}
\end{equation}

\noindent when periodic boundary conditions are applied, where $\alpha$ refers to the
coordinates axes, $x$, $y$, and $z$.

A complete theory of electric polarization in crystalline dielectrics has
been developed in recent years,
\cite{kingsmith1993a,vanderbilt1993a,souza2000a} which validates the
calculation of the dipole moments of single ions or molecules from the
center of charge of the subset of MLWF which are localized in their
vicinity.\cite{silvestrelli1999a,bernasconi2001a,bernasconi2002a} The MLWF
centers are computed according to \cite{silvestrelli1999a}
\begin{equation}
  r^w_{n,\alpha} = -\frac{L}{2\pi} \Im(\log{s_{n,\alpha}})
\end{equation}

\noindent and the induced dipole moment of a given ion $i$ is defined, in
atomic units, as
\begin{equation}
\bmu^i= -2 \sum_{n\in i} {\bm r'}^w_n
\end{equation}

\noindent where ${\bm r'}^w_n$ is measured with respect to the position of the ion. In ionic systems the attribution of a Wannier center to a given ion can be done unambiguously. Several routes have also been proposed to determine the induced quadrupole moments. Aguado {\it et al.} have exploited the fact that the MLWFs are well-localised inside the periodic cell and evaluate the components of the quadrupole on an ion $i$ from the real-space integral of the charge densities of the MLWFs on that ion:
\begin{equation}
\theta^i_{\alpha\beta}=-2\sum_{n\in i} \int_{V_{cut}^i} {\rm d}{\bf r}\mid \phi^w_n({\bf r})\mid^2 (3r'^i_\alpha r'^i_\beta -(r'^i)^2 \delta_{\alpha\beta})/2
\end{equation}
\noindent Here the integral runs over the space within a sphere of radius $r_{cut}$ around $i$, and $r'^i$ is the distance from ${\bf r}$ to the nucleus position $\bfr^i$ calculated with a minimum image convention.~\cite{aguado2003b} This method is implemented in the CASTEP code. Another method consists in determining the second moment of each Wannier function:
\begin{equation}
\langle r_\alpha r_\beta \rangle_n = -\frac{L^2}{8\pi^2}\left( \ln \frac{\mid\bra{\phi_n^w}e^{-i \frac{2\pi}{L} r_\alpha}e^{-i \frac{2\pi}{L} r_\beta}\ket{\phi_n^w}\mid^2}{\mid\bra{\phi_n^w}e^{-i \frac{2\pi}{L} r_\alpha}\ket{\phi_n^w}\mid^2\mid\bra{\phi_n^w}e^{-i \frac{2\pi}{L}  r_\beta}\ket{\phi_n^w}\mid^2} \right)
\end{equation}
\noindent The quadrupole moment of an ion is then given by~\cite{silvestrelli1999c}
\begin{equation}
\theta^i_{\alpha\beta}=-2\sum_{n\in i}\langle 3r'_\alpha r'_\beta -r'^2\delta_{\alpha\beta} \rangle_n
\end{equation}
\noindent where $\langle r'_\alpha r'_\beta\rangle_n$ is measured again with respect to the position of the ion. The advantage of this method, which is implemented in the CPMD code, is that there is no need to define a cut-off radius around $i$ (apart from the attribution of the Wannier center to the ion). Sagui {\it et al.} have extended this approach for the calculation of higher multipole moments.~\cite{sagui2004a} The generalization of the force fitting procedure to multipoles requires the minimization of the functions:
\begin{eqnarray}
\chi_D^2=\frac{1}{N_c}\sum_{j=1}^{N_c}\frac{1}{N_j}\sum_{i=1}^{N_j}\frac{\mid\bmu^i_{\rm DFT}-\bmu^i_{\rm classical}\mid^2}{\mid\bmu^i_{\rm DFT}\mid^2}\\
\chi_Q^2=\frac{1}{N_c}\sum_{j=1}^{N_c}\frac{1}{N_j} \sum_{i=1}^{N_j}\frac{\sum_{\alpha\beta}\mid\theta^i_{\alpha\beta, {\rm DFT}}-\theta^i_{\alpha\beta, {\rm classical}}\mid^2}{\sum_{\alpha\beta}\mid\theta^i_{\alpha\beta, {\rm DFT}}\mid^2}
\label{eq:chi2multipoles}
\end{eqnarray}

This fitting procedure has now been applied to an important variety of ionic systems. Here we will describe the parameterization of the oxide and fluoride series, for which the AIM and PIM frameworks were chosen respectively.

An important issue which remains is how to include the dispersion terms ($C_6^{ij}$ and $C_8^{ij}$ coefficients) in the fitting procedure. It is well-known that the dispersion interactions can not be represented accurately by a local or a semilocal exchange-correlation functional, such as that provided by the commonly used LDA and GGA functionals.~\cite{hult1999a}  We have therefore decided not to include the dispersion parameters in the fitting procedure, but rather to fix them to some given values. The choice of these values is discussed in the next section. Now the second important question is whether it is appropriate or not to include these dispersion effects when calculating the classical forces during the fitting procedure. Here our choice depends on the nature of the functional used for the DFT calculations. It is well known that the LDA functional leads to overbinding effects (although those are not due to dispersion), while GGA functionals like PBE~\cite{perdew1996a} or BLYP~\cite{becke1988a,lee1988a} lead to underbinding effects. We have therefore chosen to include the dispersion effects in the calculation of the classical forces when using the LDA functional only (the effect is included, but the corresponding parameters are kept fixed so that they are not fitted during the procedure).

\begin{figure}[ht!]
\begin{center}
\includegraphics[width=8.0cm]{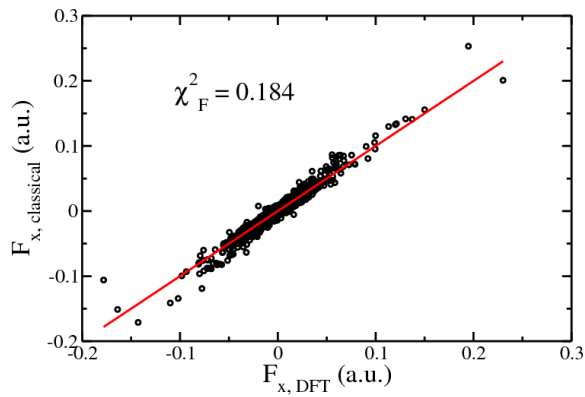}
\caption{\label{fig:chi2oxides} The quality of the fits of the $x$-component of the forces on the ions to those obtained from the {\it ab initio} calculations is shown for a series of 19 configurations (including high and low pressure polymorphs of Al$_2$O$_3$, MgO, SiO$_2$, MgAl$_2$O$_4$, MgSiO$_3$, CaO, CaAl$_2$Si$_2$O$_8$ and two Al$_2$O$_3$ melts). }
\end{center}
\end{figure}

In the case of oxide systems, our objective was to develop a potential suitable for simulation of Earth materials under crustal to lower mantle conditions of temperature and pressure.~\cite{jahn2007b} The most abundant species are those of the so-called CMAS system, i.e. the cations Ca$^{2+}$, Mg$^{2+}$, Al$^{3+}$ and Si$^{4+}$. In this work, the LDA functional has been used for the DFT calculations : A detailed comparison between GGA and LDA based potentials was undertaken for Al$_2$O$_3$; this study showed that the description of the melt was much better for the latter.~\cite{jahn2006a} The AIM functional form was used for the classical interaction potential. We can therefore summarize the fitting procedure this way:
\begin{enumerate}
\item Generation of a series of typical condensed-phase configurations
\item DFT calculations on each of these configurations:
\begin{enumerate}
\item Determination of the ground-state wavefunctions, which gives access to the {\it ab initio} forces and stress tensor components
\item Wannier localization, from which the {\it ab initio} induced dipoles and quadrupoles components are calculated
\end{enumerate}
\item Determination of the dispersion coefficients (see next section)
\item Minimization of $\chi_D$ and $\chi_Q$ with respect to the parameters of the polarization term ($V^{\rm polarization}$)
\item Minimization of $\chi_F$ and $\chi_S$ with respect to the parameters of the repulsion term ($V^{\rm repulsion}$)~\footnote{ Note that all the $\chi_i$ values could be minimized together but the use of a two-step procedure permits to avoid some cancellation of errors in the parameterization.}
\end{enumerate}

The quality of the fits of the $x$-component of the forces on the ions to those obtained from the {\it ab initio} calculations is shown for a series of 19 configurations  of approximately 100 atoms (including high and low pressure polymorphs of Al$_2$O$_3$, MgO, SiO$_2$, MgAl$_2$O$_4$, MgSiO$_3$, CaO, CaAl$_2$Si$_2$O$_8$ and two Al$_2$O$_3$ melt) on figure \ref{fig:chi2oxides}, and the parameters obtained for the CMAS system with this procedure are summarized in table \ref{tab:aimcmas}.

\begin {table}[t]
\begin {center}
\begin {tabular} { c|c|c|c|c|c }
\hline
 Ion pair       & O$^{2-}$-O$^{2-}$   & Ca$^{2+}$-O$^{2-}$  & Mg$^{2+}$-O$^{2-}$  & Al$^{3+}$-O$^{2-}$  & Si$^{4+}$-O$^{2-}$  \\
\hline
$A^{ij}$       & 1068.0& 40.168& 41.439& 18.149& 43.277\\
$a^{ij}$       & 2.6658& 1.5029& 1.6588& 1.4101& 1.5418\\
$B^{ij}$       &       & 50532.& 59375.& 51319.& 43962.\\
$b^{ij}$       &       & 3.5070& 3.9114& 3.8406& 3.9812\\
$C^{ij}$       &       & 6283.5& 6283.5& 6283.5& 6283.5\\
$c^{ij}$       &       & 4.2435& 4.2435& 4.2435& 4.2435\\
\hline
$b_{D}^{ij}=b_{D}^{ji}$   &       & 2.0261& 2.2148& 2.2886& 2.1250\\
$c_{D}^{ij}$   &       & 3.9994& 2.8280& 2.3836& 1.5933\\
$b_{Q}^{ij}$   &       & 1.5297& 1.9300& 2.1318& 1.9566\\
$c_{Q}^{ij}$   &       & 1.6301& 1.3317& 1.2508& 1.0592\\
\hline
$C_6^{ij}$     & 44.372& 2.1793& 2.1793& 2.1793& 2.1793\\
$C_8^{ij}$     & 853.29& 25.305& 25.305& 25.305& 25.305\\
$b_{6}^{ij}=b_{8}^{ij}$& 1.4385& 2.2057& 2.2057& 2.2057& 2.2057\\
\hline
\hline
$D$            &0.49566&       &$\beta$& 1.2325&\\
$\zeta$        &0.89219&       & $\eta$& 4.3646&\\
$\alpha$       & 8.7671&       &    $C$&11.5124&\\
\hline
\end {tabular}
\caption{Parameters in the repulsive and polarization parts of the potential for the CMAS system.
All values are in atomic units. The dipole-dipole-quadrupole polarizability was set to 0.}
\label{tab:aimcmas}
\end {center}
\end {table}

\begin{figure}[ht!]
\begin{center}
\includegraphics[width=8.0cm]{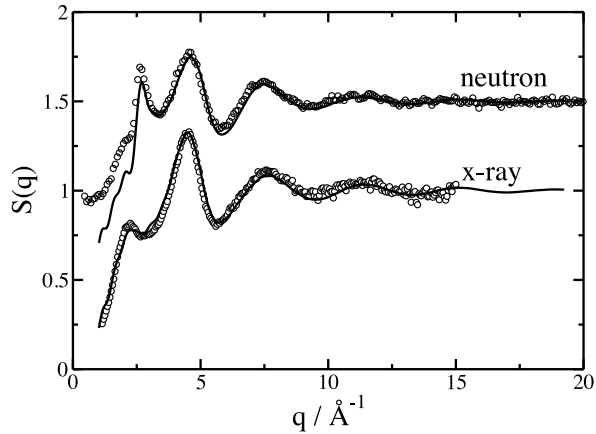}
\caption{\label{fig:xrayneutron} { Total static structure factors of MgAl$_2$O$_4$ at $T$~=~2500~K from MD simulation~\cite{jahn2008a} (lines) compared to experimental neutron and x-ray diffraction data (circles) at $T$~=~2423~K.~\cite{hennet2007a} The neutron structure factors are offset by a value of 0.5.} }
\end{center}
\end{figure}

The potential reproduces well lattice parameters and elastic constants of
various oxides and silicates of the CMAS system. Typically, the lattice
constants from the simulations are up to a few percent lower than the 
experimental values, which is consistent with the use of LDA for the DFT 
reference calculations. The model also provides reasonable predictions for the 
thermal expansion and the volume compression of major phases of the Earth's 
mantle in a wide pressure range.~\cite{jahn2007b}. Liquid state densities of
oxide and silicate melts are usually within the error bars of the experimental
data. The model performs somewhat less well for open network-forming
structures close to the pure SiO$_2$ composition.

Applications of the
potential include the investigation of structural and physical properties of 
solid and liquid oxides relevant in geological or technological context. 
A number of studies were concerned with the stability and mechanisms of phase 
transitions in magnesium silicates (see \cite{jahn2010a} for a summary). 
The atomic structure and dynamics of magnesium and calcium aluminate liquids 
was studied in combination with neutron and X-ray scattering experiments using 
aerodynamic levitation techniques.~\cite{krishnan2005a,jahn2007a,pozdnyakova2007a,drewitt2011a}  An example of comparison between simulated~\cite{jahn2008a} and experimental~\cite{hennet2007a} total structure factors of liquid MgAl$_2$O$_4$ is shown on figure \ref{fig:xrayneutron}.
The almost quantitative agreement between measured and computed static and
dynamic structure factors underlines the high accuracy and transferability of 
the potential, which was optimized using mostly solid state configurations. 
The structural studies illustrate
that molecular simulations are essential to interpret the diffraction data of 
multicomponent liquids since even two different experiments (neutron and X-ray
diffraction) do not allow the unambiguous determination of some of the
anion-cation nearest neighbor distances and cation coordination 
numbers.~\cite{drewitt2011a}
Besides the structural model, the simulations provide insight into the
relation between melt structure and physical properties, such as the viscosity
or the self-diffusivities.~\cite{jahn2006a,jahn2008a} 
The latter are important parameters,
e.g., to understand the evolution of the early Earth, which may have been 
covered by a magma ocean composed of silicate liquid. Recent simulations of
Mg$_2$SiO$_4$ liquids using the CMAS potential give new constraints on the
thermodynamic and rheological properties of the melt in the relevant range of 
pressure and temperature.~\cite{adjaoud2008a,adjaoudsubmitted} A number of
ongoing projects make use of the CMAS potential to study solid-solid and 
solid-liquid interfaces, the structure of complex silicate melts in
conjunction with X-ray absorption spectroscopy or with the interpretation of 
trace element partioning data between solid and liquid silicates.

\begin{figure}[ht!]
\begin{center}
\includegraphics[width=8.0cm]{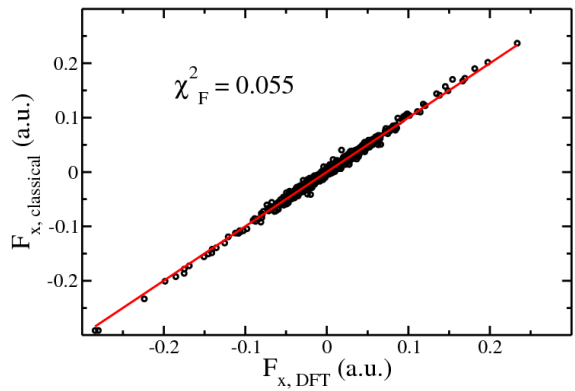}
\caption{\label{fig:chi2fluorides} The quality of the fits of the $x$-component of the forces on the ions to those obtained from the {\it ab initio} calculations is shown for a series of 25 configurations (pure LiF, NaF, KF, CaF$_2$, SrF$_2$, YF$_3$, LaF$_3$ and ZrF$_4$, LiF-YF$_3$ and LiF-LaF$_3$ mixtures). }
\end{center}
\end{figure}

\begin{table}
\begin{center}
\begin{tabular}{c | c }
\hline
Ion type & Polarizability ($\alpha^i$) \\
\hline
F$^-$ & 7.9  \\
Li$^+$ & 0 \\
Na$^+$ & 1.0 \\
K$^+$ & 5.0 \\
Rb$^+$ & 8.4 \\
Cs$^+$ & 14.8 \\
Ca$^{2+}$ & 3.1 \\
Sr$^{2+}$ & 5.1 \\
Y$^{3+}$ & 3.8 \\
La$^{3+}$ & 7.5 \\
Zr$^{4+}$ & 2.9 \\
\hline
\end{tabular}
\end{center}
\caption{\label{tab:param-polarisation} Fitted polarizabilities for the various ions in molten fluorides. All values are in atomic units.}
\end{table}

\begin{table}
\begin{center}
\begin{tabular}{l | c | c| c| c | c }
\hline
Ion pair & $A^{ij}$ & $a^{ij}$ & $b^{ij}_D=b^{ji}_D$ & c$^{ij}_D$ &  c$^{ji}_D$ \\
\hline
F$^-$-F$^-$ & 282.3 & 2.444 & -- & 0.0 & 0.0 \\
F$^-$-Li$^+$ & 18.8 & 1.974 & 1.834 & 1.335 & \\
F$^-$-Na$^+$ & 44.9 & 1.927 & 1.831 & 2.5 & 0.022\\
F$^-$-K$^+$ & 138.8 & 2.043 & 1.745 & 2.5 & -0.31 \\
F$^-$-Rb$^+$ & 151.0 & 1.961 & 1.822 & 3.5 & -0.44 \\
F$^-$-Cs$^+$ & 151.1 & 1.874 & 1.930 & 3.4 &  0.49 \\
F$^-$-Ca$^{2+}$& 73.0 & 1.859 & 1.732 & 1.5 & -0.31 \\
F$^-$-Sr$^{2+}$& 82.5 & 1.778 & 1.554 & 1.331 & -0.33\\
F$^-$-Y$^{3+}$& 87.4 & 1.832 & 1.847 & 1.966 & -0.89\\
F$^-$-La$^{3+}$& 161.6 & 1.867 & 1.614 & 1.348 &-0.47 \\
F$^-$-Zr$^{4+}$& 62.6 & 1.74 & 1.882 & 1.886& -1.0 \\
$M^{m+}$-$M^{m+}$& 1.0 & 5.0 & -- & 0.0 & 0.0 \\
$N^{+}$-$N^{+}$& 5000.0  & 3.0 & -- & 0.0 & 0.0 \\
\hline
\end{tabular}
\end{center}
\caption{\label{tab:param-repulsion} Fitted parameters for the repulsion and polarization terms for molten fluorides. $b_6^{ij}=b_8^{ij}=1.9$ for all the pairs. $M^{m+}$ stands for any cation between Li$^+$, Na$^+$, K$^+$, Ca$^{2+}$, Sr$^{2+}$, Y$^{3+}$, La$^{3+}$ or Zr$^{4+}$, while $N^+$ is Rb$^+$ or Cs$^+$. All values are in atomic units.}
\end{table}

In the case of fluoride systems, we are mainly interested in determining physico-chemical properties of melts including many different components. We have therefore built a transferable PIM interaction potential which includes Li$^+$, Na$^+$, K$^+$, Rb$^+$, Cs$^+$, Be$^{2+}$, Ca$^{2+}$, Sr$^{2+}$, Y$^{3+}$, La$^{3+}$ and Zr$^{4+}$ ions. In this work, the PBE functional has been used for the DFT calculations (unlike the case of oxide materials, no test has been made using the LDA). We can therefore summarize the fitting procedure this way:
\begin{enumerate}
\item Generation of a series of typical condensed-phase configurations
\item DFT calculations on each of these configurations:
\begin{enumerate}
\item Determination of the ground-state wavefunctions, which gives access to the {\it ab initio} forces  components
\item Wannier localization, from which the {\it ab initio} induced dipoles components are calculated
\end{enumerate}
\item Minimization of $\chi_D$  with respect to the parameters of the polarization term ($V^{\rm polarization}$)
\item Minimization of $\chi_F$  with respect to the parameters of the repulsion term ($V^{\rm repulsion}$)
\item Determination of the dispersion coefficients (see next section)
\end{enumerate}

Small differences are observed with the oxide materials situation: Due to the smaller number of parameters involved in the fitting procedure and the absence of quadrupole polarization effects in the PIM, only $\chi_D$ and $\chi_F$ had to be minimized in this case. In addition the dispersion coefficients can be determined after the fitting step because the dispersion term is set to 0 when fitting the repulsion (due to the use of a GGA functional in the DFT calculations). Since we are mainly interested in systems which are in the liquid state, this procedure was in fact performed twice: Firstly, with a series of crystalline configurations, which led to a first set of parameters. And secondly with a series of liquid state configurations  of approximately 100 atoms generated from molecular dynamics simulations with the parameters of the first step. The quality of the fits of the $x$-component of the forces on the ions to those obtained from the {\it ab initio} calculations is shown for a series of 25 configurations (pure LiF, NaF, KF, CaF$_2$, SrF$_2$, YF$_3$, LaF$_3$ and ZrF$_4$, LiF-YF$_3$ and LiF-LaF$_3$ mixtures) on figure \ref{fig:chi2fluorides}, and the parameters obtained for the fluoride system with this procedure are summarized in tables \ref{tab:param-polarisation} and \ref{tab:param-repulsion}.

\begin{figure}[ht!]
\begin{center}
\includegraphics[width=8.0cm]{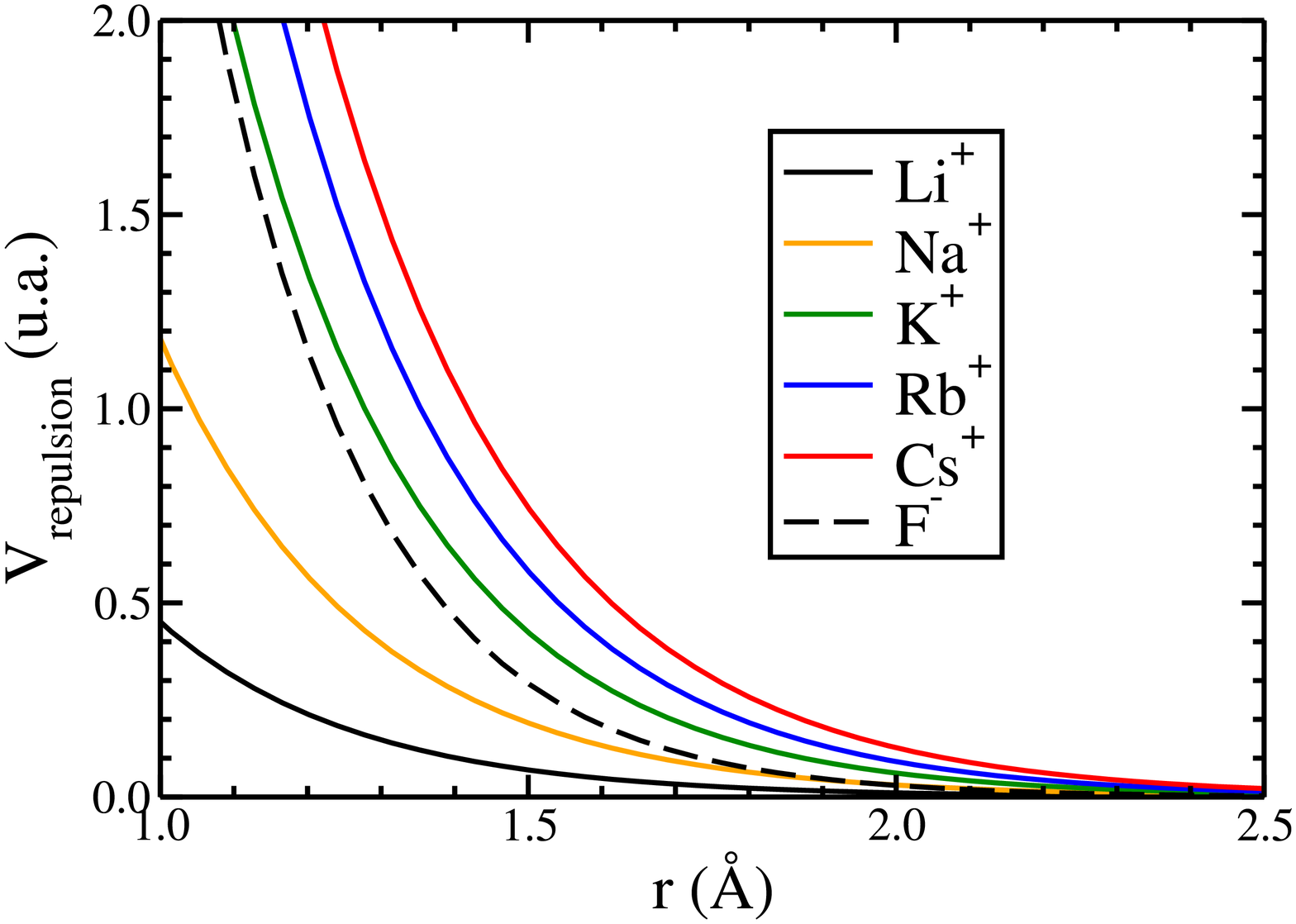}
\includegraphics[width=8.0cm]{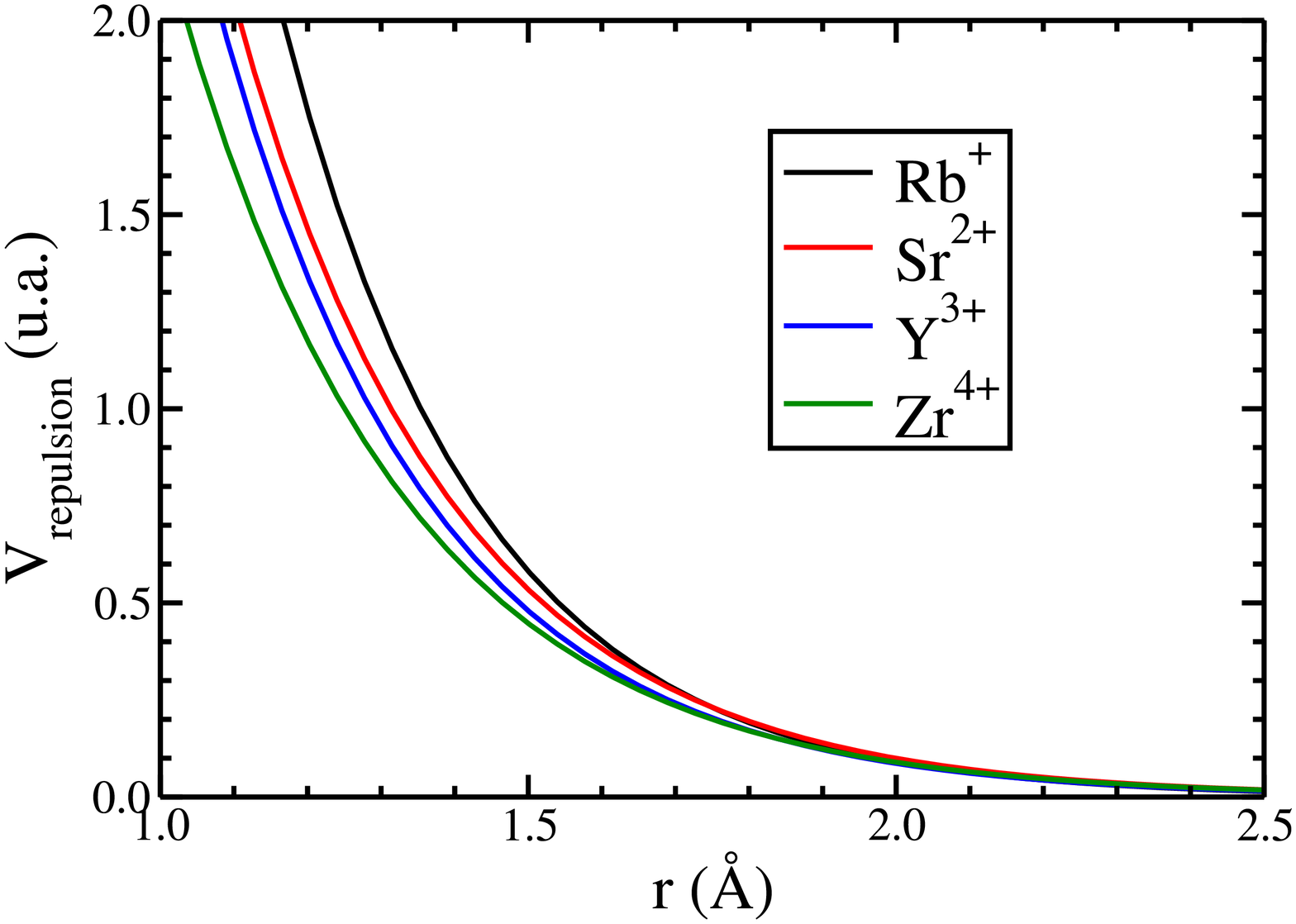}
\caption{\label{fig:comparaison-param} Comparison of the repulsion terms obtained for various cation-fluoride pairs. Top: Alkaline cations Li$^+$, Na$^+$, K$^+$, Rb$^+$ and Cs$^+$; bottom: period 5 elements Rb$^+$, Sr$^{2+}$, Y$^{3+}$ and Zr$^{4+}$}
\end{center}
\end{figure}

A first test of the potential is provided by a simple comparison of the parameters within a series of chemical species. For example, in figure \ref{fig:comparaison-param} we compare the repulsive term of the potential for various M$^{m+}$-F$^-$ ion pairs. Firstly, in the case of the alkaline cations, we observe that the repulsion wall is pushed towards longer distances in a way which is coherent with the increase of the ionic radius of the cation in this series. More interesting is the second test, which compares the period 5 elements Rb$^+$, Sr$^{2+}$, Y$^{3+}$ and Zr$^{4+}$. All these ions have the same number of electrons and their ionic radius therefore do not differ a lot. The repulsion walls are very close one from each other and they become shorter-ranged when the cationic charge is increased, which is in agreement with chemical intuition.

\begin{figure}[ht!]
\begin{center}
\includegraphics[width=8.0cm]{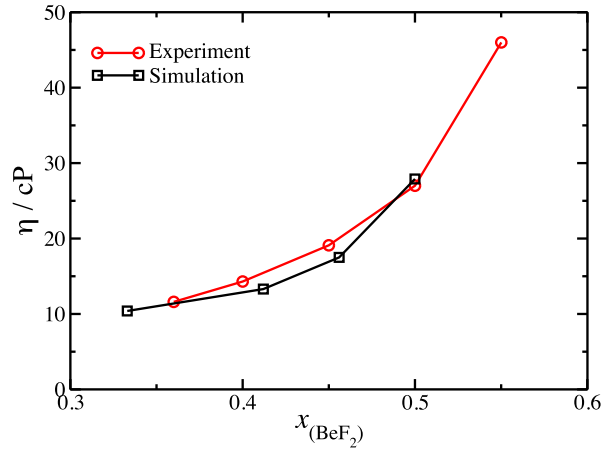}
\caption{\label{fig:viscolifbef2} { Comparison of the calculated~\cite{salanne2006a} and experimental~\cite{cantor1969a} viscosities of LiF-BeF$_2$ mixtures with varying compositions at $T$~=~873~K.} }
\end{center}
\end{figure}

The interest in molten fluorides is mainly due to their involvement in numerous current (aluminum production) and proposed (molten salt reactors) technologies. They are often dangerous and corrosive materials, which are liquid at high temperatures, so that performing systematic experimental studies is difficult. The development of accurate models which are able to predict the physical properties over a wide range of conditions  is therefore very valuable. Among all the molten fluorides, an enormous amount of experimental work has been done on LiF-BeF$_2$ mixtures and it is by far the best-characterized fluoride melt.~\cite{cantor1969a,vaslow1973a} For this reason, it was the ideal testing ground for evaluating our interaction potential development approach and we focused first on this system.~\cite{heaton2006a} We could show that our model performed very well in reproducing the polymeric fluoroberyllate speciation and the vibrational properties in these melts.~\cite{salanne2006a} Calculated dynamic properties (electrical conductivity, viscosity) also agree quantitatively with the experimental data:~\cite{salanne2007b}  To illustrate this, we show a comparison of the calculated~\cite{salanne2006a} and experimental~\cite{cantor1969a} viscosities of LiF-BeF$_2$ mixtures with varying compositions at $T$~=~873~K on figure \ref{fig:viscolifbef2}. The interfacial properties were also tested by calculating the surface tension from a simulation of the liquid-vapor interface.~\cite{salanne2007a} In a second step, a series of simulation studies  was carried on for other molten fluoride systems. This allowed us to predict heat-transport properties of LiF-NaF-KF and NaF-ZrF$_4$ mixtures~\cite{salanne2009a} in order to evaluate their suitability to serve as primary coolant in the advanced high-temperature reactor concept.~\cite{forsberg2005a} Our simulations were also used to interpret EXAFS and NMR experimental data on fluorozirconate melts~\cite{pauvert2010a} and to calculate electrical conductivities~\cite{salanne2009b}, diffusion coefficients~\cite{saroukanian2009a} and internal mobilities~\cite{merlet2010a}.

\section{Beyond force-matching: direct calculation of parameters}

Although the fitting procedure presented in the previous section has proven to be very successful, in particular in the context of the development of transferable potentials, it can sometimes be enhanced through the direct {\it calculation} of some of the parameters involved in the models. Among all the parameters, the dipolar polarizability $\alpha$ has an evident physical meaning since it measures the relative tendency of the electron cloud of the ion to be distorted from its normal shape by an electric field. Although one could think that the polarizability of a molecule or an ion is an intrinsic property, which remains unchanged from the gas phase to any condensed phase, this is not the case due to the effect of the confining environmental potential in the latter \cite{fowler1984a,fowler1985a}.

Again, in order to determine the condensed phase polarizabilities from a direct DFT calculation, the MLWF formalism provides a very convenient route.~\cite{heaton2006b}  When a small external electric field $\bm{\mathcal{E}}$ is
applied to the system, the linear response may be characterized by an
additional field-induced dipole moment $\bdelta\bmu^i$ on each individual
ion. It is convenient to think of the applied field as an optical
field, in order to distinguish its effect from that of the static fields,
which are caused by the permanent charge distributions of the molecules,
and to think of $\bdelta\bmu^i$ as the net induced dipole which is
oscillating at the optical frequency. For an electronically insulating
material, the induced dipole can be written in terms of the total (optical
frequency) electric field which acts on it,
\begin{equation}
\bdelta\bmu^i=\balpha^i  \cdot
\left[\bm{\mathcal{E}} + \sum_{j\ne i}{\bm T}^{(2)}\cdot
\bdelta\bmu^j \right] \label{eq:defpol}
\end{equation}
where the sum runs over all polarizable ions $j\ne
i$ in the system.  In this equation, we have introduced the dipole
polarizability tensor $\balpha^i$ of ion $i$, for a
particular condensed phase configuration. Also included is
the dipole-dipole interaction tensor, ${\bm T}^{(2)}$, whose components
are defined by $T^{(2)}_{\alpha\beta}=\bnabla_\alpha \bnabla_\beta
\frac{1}{r^{ij}}$; in practice, for a periodic system, it will be
computed using the Ewald summation
technique.\cite{aguado2003a,laino2008a} The first term on the right-hand
side of the equation represents the direct contribution of the external
electric field to the induced dipole; the second-term is the contribution
of the reradiated electric fields due to the dipoles which are induced in
all the other ions ($j\ne i$) in the sample. In principle, higher
order induced multipoles also contribute to this expansion, but we will
ignore them. In a uniform external field, the directly induced higher order
multipoles on spherical atoms and ions vanish, and even for molecules,
their effect is expected to be much smaller than that of the dipoles.

In DFT calculations on periodic systems, the coupling between the external
electric field and the electronic system is expressed through the
macroscopic polarization of the periodically replicated cell
\cite{umari2002a,umari2005a} and is defined using the Berry phase approach
of Resta.\cite{resta1998a} It is then possible to determine the new
partial dipole moment for each species in the presence of a field, {\it
  via} another localization step. The field induced dipoles are calculated
from the difference between the total molecular dipoles in the presence and
absence of the field.

Equation \ref{eq:defpol} can be inverted to determine the individual
electronic polarizabilities for that particular condensed-phase
configuration. Consider the application of fields
$\bm{\mathcal{E}}^{(\alpha)}$, along each Cartesian direction
$\alpha=x,y,z$, and denote by $\{\bdelta\bmu^{i,(\alpha)}\}$
the corresponding values of the induced dipole moments. The total field
$\bm{f}^{i,(\alpha)}$ at each position $\bm{r}^i$ can be obtained from
\begin{equation} {\bm f}^{i,(\alpha)}=\bm{\mathcal{E}}^{(\alpha)} +
  \sum_{j\ne i} {\bm T}^{(2)} \cdot \bdelta\bmu^{j,(\alpha)}
\end{equation}
which is conveniently evaluated from the electric field given by a dipolar
Ewald sum. Finally, the polarizability tensor of ion $i$ is given by
\begin{equation}
\balpha^i=({\bm F}^i)^{-1}\cdot \bPi^i
\label{eq:polawoff}
\end{equation}
\noindent where ${\bm F}^i$ and $\bPi^i$ are second-rank
  three-dimensional tensors defined as
\begin{eqnarray}
F^i_{\alpha\beta}=f^{i,(\beta)}_\alpha\\
\Pi^i_{\alpha\beta}=\delta\mu^{i,(\beta)}_\alpha
\end{eqnarray}

The results obtained for a series of molten fluorides and chlorides are summarized in table \ref{tab:calcpolarizabilities}. Important environmental effects occur when the nature of the liquid is changed. In the fluoride melts, for example, the polarizability of the fluoride anion
shifts a lot, passing from 7.8~a.u. in molten LiF (which is very close to the value obtained from the force-fitting procedure, this is also the case for the cation polarizabilities)
 to 11.8~a.u. in CsF. This is due to differences in the confining potential, which affects
the electron density around a given anion, and  originates from both Coulombic
interactions and the exclusion of electrons from the region occupied by the
electron density of the first-neighbor shell of cations.~\cite{fowler1984a,fowler1985a,jemmer1998a} When
passing from one cation to another (for example in the series
Li$^+$$\rightarrow$Na$^+$$\rightarrow$K$^+$$\rightarrow$Cs$^+$), two effects are
then competing: On the one hand, the anion-cation distance increases, which results
in a diminution of the confining potential, but on the other hand, the volume
occupied by the cation electron density also increases, with an opposite effect on
the confining potential. Here, the observed increase of polarizability with the size
of the cation tends to show that the first effect is the most important. Indeed,
the value obtained for CsF is approaching the free F$^-$ anion polarizability,
which is 16~a.u. These effects have also been observed in similar calculations performed on solid oxides~\cite{heaton2006b} and protic solvents~\cite{buin2009a,salanne2011b}. They have indirectly been used for building quantitative Lewis acidity scales in inorganic materials.~\cite{duffy1971a,angell2009a} Such a result means that in order to build a completely transferable interaction potential, the model should  allow the variation of the polarizability in response to the fluctuations of the environments. First steps have been made in that direction in the case of MgO only.~\cite{aguado2004a,aguado2005a}

\begin{table}[!ht]
\begin{center}
\begin{tabular}{lcc}
\hline
System & Ion type & Polarizability ($\alpha^i$ )\\
\hline
LiF & F$^-$ &  7.8  \\
    & Li$^+$ & 0.3  \\
NaF & F$^-$ & 9.6   \\
    & Na$^+$ & 1.1 \\
KF & F$^-$& 10.7 \\
   & K$^+$& 5.5 \\
CsF & F$^-$& 11.8 \\
   & Cs$^+$& 16.3 \\
\hline
NaCl & Cl$^-$ & 25.1\\
     & Na$^+$ & 1.2\\
KCl & Cl$^-$ & 27.0  \\
    & K$^+$ & 5.5 \\
\hline
\end{tabular}
\end{center}
\caption{ \label{tab:calcpolarizabilities}
Calculated ionic polarizabilities in the liquid phase. All values are in atomic units.
}
\end{table}

Since the centers of the MLWFs allow one to derive atomic dipoles in the condensed
phase and hence atomic polarizabilities, it is tempting to exploit further
these localized orbitals, which are often interpreted in terms of non-bonding
orbitals or lone pairs, to derive the remaining terms of the force field.
A systematic procedure has been introduced recently for this purpose,
in which a series of approximations leads to the attractive and repulsive van der
Waals interactions from the knowledge of the geometric properties of the
MLWF~\cite{rotenberg2010a}.

\begin{figure}[ht!]
\begin{center}
\includegraphics[width=4cm]{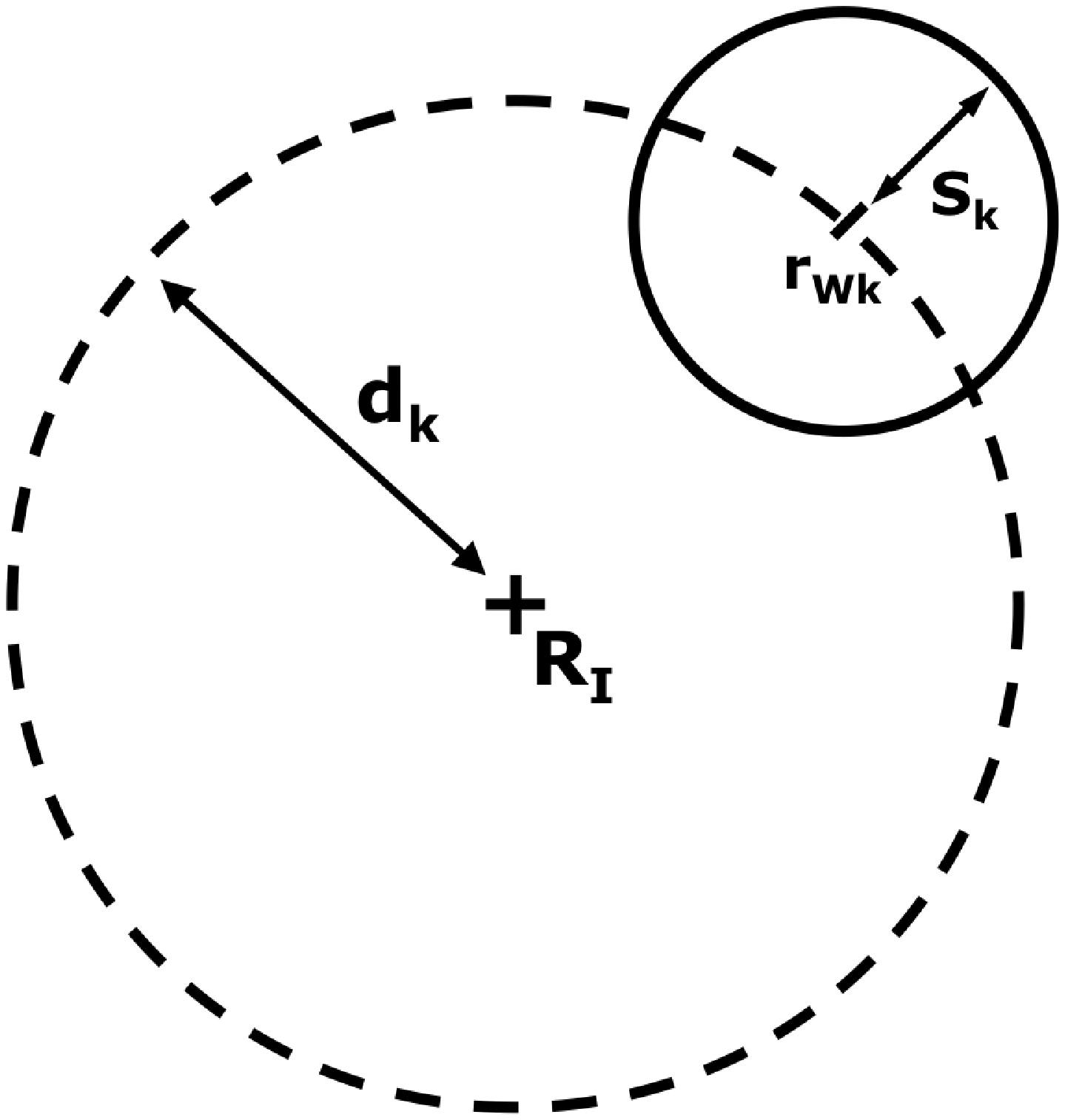}
\includegraphics[width=8cm]{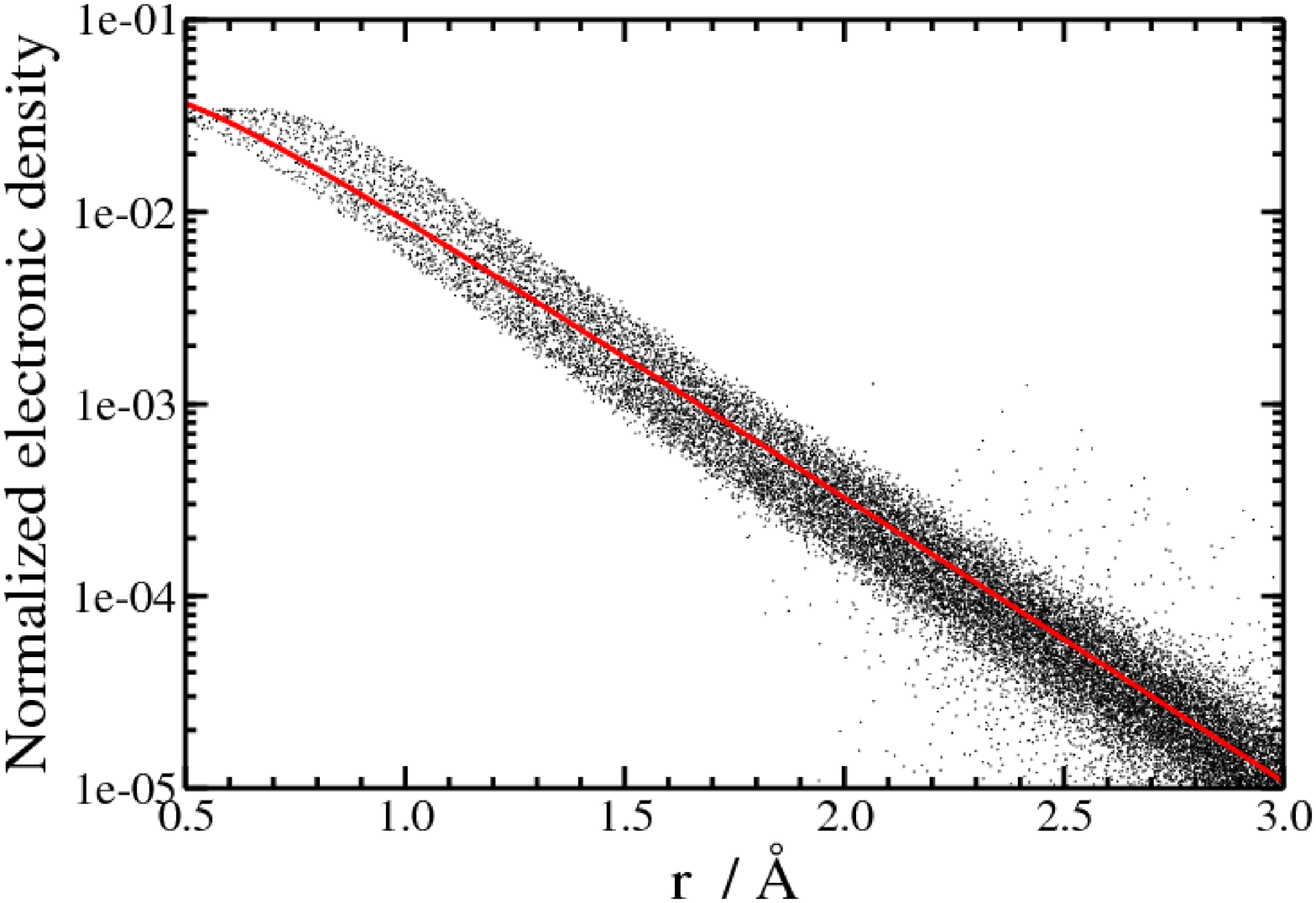}
\end{center}
\caption{\label{fig:electronicdensity}
Electronic density around Cl$^-$ ions in molten KCl
from DFT (scattered plot), compared to the analytic ansatz (equation \ref{eq:rho8e}).
which involves only the spread and distance to nucleus of the MLWFs.
}
\end{figure}

The central idea of this approach consists of assigning
a localized electronic density considered as frozen around each nucleus,
reconstructed from the associated MLWFs.
The exponential decay of MLWFs in electronic insulators~\cite{brouder2007a}
suggests modeling them as Slater orbitals determined solely by
their spread $S_k$
and center located at $\bfr^w_k$ determined from the localization
procedure~\cite{silvestrelli2008a}. These two properties can be obtained
for each type of MLWF from averages over liquid configurations.
Each MLWF contributes to an electronic density
$\rho_{Wk}=\displaystyle
n_k\frac{\kappa_k^3}{8\pi}e^{-\kappa_k\absv{\bfr-\bfr^w_k}}$
with $n_k=2$ the number of electrons in the orbital and
$\kappa_k=\frac{2\sqrt{3}}{S_k}$.
In order to obtain orientationally averaged interaction potentials,
one further proceeds to an orientational average of the density around each
nucleus $i$ (see figure \ref{fig:electronicdensity}),
assuming an isotropic distribution of the $N_W$
MLWFs centers with the result:
\begin{equation}
\label{eq:rho8e}
\rho^i(\bfr) = \sum_{k=1}^{N_W}
    \frac{n_k\kappa_k}{8\pi\rsup\rinf} e^{\displaystyle-\kappa_k \rsup}
    \braa (1+\kappa_k\rsup) \sinh(\kappa_k \rinf)
   - \kappa_k\rinf \cosh(\kappa_k\rinf) \kett
\label{eq:densitywoff}
\end{equation}
where $\rsup=\max(\absv{\bfr-\bfr^w_k},d_k)$ and
$\rinf=\min(\absv{\bfr-\bfr^w_k},d_k)$, with $d_k=\absv{\bfr^w_k-\bfr^i}$
the distance of the $k$th MLWF center to the nucleus.
The spread and distance to nucleus of Wannier orbitals for molten
NaF, NaCl, KF and KCl are summarized in table~\ref{tab:WOproperties}.

Figure~\ref{fig:electronicdensity} also compares the electronic density around Cl
nuclei with the spherically averaged result given by equation \ref{eq:rho8e}.
The agreement with the mean density is very good even
at short distances, the scattering of the measured electronic densities
being mainly due to thermal fluctuations not reproduced by the
rigid density approximation.
Exponential decay of the molecular density is also
apparent over more than two decades and valid in the important region where
the electronic densities of neighbouring ions overlap.
For one of the considered species (K$^+$ in molten KF and KCl),
the electronic density around the nucleus, derived from the MLWFs,
is better described at large distances by a nucleus-centered
Slater orbital (spread 1.30 a.u. in KF, 1.34 a.u. in KCl) than
by equation \ref{eq:rho8e}. Equation \ref{eq:rho8e} provides
a more accurate description near the nucleus, but not further away.

\begin{table}[!ht]
\begin{center}
\begin{tabular}{lccc}
\hline
System & Ion type & $\langle S^i\rangle$ & $\langle d^i\rangle$ \\
\hline
KF   & F$^-$  & 1.35 & 0.555 \\
     & K$^+$  & 1.36 & 0.685 \\
NaF  & F$^-$  & 1.32 & 0.560 \\
     & Na$^+$ & 0.85 & 0.410 \\
KCl  & Cl$^-$ & 1.92 & 0.869 \\
     & K$^+$  & 1.36 & 0.680 \\
NaCl & Cl$^-$ & 1.90 & 0.877 \\
     & Na$^+$ & 0.86 & 0.414 \\
\hline
\end{tabular}
\end{center}
\caption{ \label{tab:WOproperties}
Properties of the Wannier orbitals:
Average spread $S^i$ and distance $d^i$ to the nucleus. All values are in atomic units.
}
\end{table}

Once the localized density around each nucleus in condensed phase is known,
the repulsive van der Waals interaction, which contains a kinetic and
an exchange-correlation part, can be derived using two
approximations: The above-mentioned frozen density ansatz and
the use of a kinetic energy functional. The repulsion energy $V^{\rm repulsion}$
is computed at the DFT level from the superposition of the rigid electronic
clouds around each atom~\cite{gordon1972a,barker2003a}
and the resulting energy surface parametrized as a function
of atomic coordinates. For each interatomic separation $r$,
we compute
\begin{equation}
V^{\rm repulsion}(r) = \Ekin[\rho_s,\nabla\rho_s]
  +\int \epsilon_{xc}\left[ \rho_s,\nabla\rho_s\right] \rho_s \ {\rm d}^3\bfr,
\label{eq:repwoff}
\end{equation}
where $\rho_s=\rho^i+\rho^j$ is the superposition of frozen atomic densities which have their centers separated by a distance $r$,
$\Ekin$ and $\epsilon_{xc}$ are the kinetic
energy and exchange correlation functionals, respectively.
Any form can be used for the kinetic energy functional $\Ekin$ and
the exchange-correlation energy density $\epsilon_{xc}$,
provided that the latter is the one used in the starting DFT calculations.
However the resulting $V^{\rm repulsion}$ depends on their choice.
The LLP kinetic energy functional~\cite{lee1991a}
is generally the most accurate~\cite{barker2003a,tran2002a}.
Note that improvements of the functionals would
automatically transfer to the resulting force fields.
The resulting $V^{\rm repulsion}$ generally decays as $Ae^{-a r}$ in the region
corresponding to the first neighbour shell, thus justifying the
analytical form of the repulsion term in the PIM and the AIM and providing the associated parameters.

\begin{figure}
\begin{center}
\includegraphics[width=8.0cm]{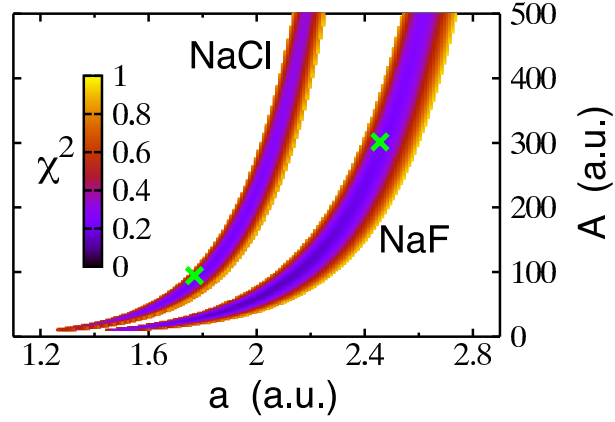}
\end{center}
\caption{\label{fig:map}
Mean-square relative error $\chi_F^2$ on the atomic forces
compared to DFT (PBE functional) in the NaCl and NaF melts,
as a function of the Na$^+$-Cl$^-$/F$^-$ repulsion parameters $Ae^{-ar}$.
The set obtained by our method ($\times$) stands well
within the $\chi_F^2<1.0$ isosurface in both cases.
}
\end{figure}

As a first example, we applied this strategy to the case of NaF and NaCl melts, starting from
DFT calculations with the PBE functional and using the LLP kinetic
functional to compute the kinetic energy in equation~\ref{eq:repwoff}.
The relevance of this approach can be appreciated on figure~\ref{fig:map},
which reports the mean-square relative error on the forces
$\chi_F^2$ defined in the previous section
as a function of the Na$^+$-Cl$^-$/F$^-$ repulsion parameters
$Ae^{-ar}$, all other parameters of the force field remaining fixed. The sets obtained by this method stand well within the
narrow windows $\chi_F^2<1.0$, thus justifying \textit{a posteriori} the frozen
density approximation for these systems. The corresponding $\chi_F^2$ are
0.241 and 0.363 for NaF and NaCl, respectively. In the latter, less favourable
case, the error is not much larger than the value obtained with a force field of the same
form (PIM) but with all parameters numerically fitted by force-matching ($\chi_F^2=0.118$),
and significantly smaller than the one obtained with the
popular Tosi-Fumi potential ($\chi_F^2=1.250$)~\cite{ohtori2009a}.

Finally, we now turn to the problem of the dispersion
interaction. As mentioned previously dispersion forces are not easily captured at the DFT level.
While the use of non-local functionals seems to be a promising approach,~\cite{zhu2010a}
their computational cost has so far limited their use in AIMD simulations,
for which empirical corrections such as the one developed by Grimme~\cite{grimme2004a}
are preferred. A promising alternative based on a simpler non-local van der Waals density functional has recently been proposed.~\cite{vydrov2010a} In the work concerning the oxide and fluoride series we have used $C_6^{ij}$ and  $C_8^{ij}$ values obtained from a limited set of {\it ab initio} (coupled-cluster or M{\o}ller-Plesset) values on in-crystal ions~\cite{fowler1985b} and mixing rules to scale values from one material to another. It is also possible to derive systematically approximate
parameters for the dispersion term defined in equation ~\ref{eq:dispersion}.
Silvestrelli showed that MLWF provide again a convenient framework to
tackle the dispersion effects and introduced the computation
{\it a posteriori} of the $-C_6^{kl}/r_{kl}^6$ interaction between Wannier
centers~\cite{silvestrelli2008a}, where $C_6^{kl}$ depends only on the
spread of the MLWFs:
\begin{equation}
C_6^{kl} = \frac{3}{32\pi^{3/2}}
     \int_{|\bfr|\leq r_c} {\rm d}\bfr
     \int_{|\bfr'|\leq r_c'} {\rm d}\bfr'
     \frac{\sqrt{\rho_{Wk}(r)\rho_{Wl}(r')}}
          {\sqrt{\rho_{Wk}(r)}+\sqrt{\rho_{Wl}(r')}}
\label{eq:dispwoff1}
\end{equation}
where the cut-off radius for each orbital depends on its spread
as
\begin{equation}
r_c=\left[ 1.475-0.866\ln S \right] S.
\end{equation}
\noindent  These coefficients are obtained from
the MLWFs using the expression proposed by Andersson \textit{et al.}~\cite{andersson1996a}
for the long-range interaction between two separated fragments.
One can then obtain the dispersion interaction between a pair of atoms
as the averaged sum over pairs of MLWFs
(for $k$, $l$ from different sites)~\cite{rotenberg2010a}.
Assuming an isotropic distribution of centers around
the nuclei $i,j$ at fixed distance $d_{k,l}$,
we obtain to 2$^{nd}$ leading order:
$V^{\rm dispersion} = - \sum_{n=6,8}C_n^{ij}/(r^{ij})^n$
where the dispersion coefficients are:
\begin{eqnarray}
C_6^{ij} &=& \sum_{k\in i,l\in j} C_6^{kl} \\
C_8^{ij} &=& \sum_{k\in i,l\in j} 5(d_k^2+d_l^2)C_6^{kl}
\label{eq:dispwoff2}
\end{eqnarray}
We recall that the distances $d_{k,l}$ result from the orbital localization procedure
and are not adjustable parameters. This approach thus provides two of the parameters
entering in equation~\ref{eq:dispersion}, leaving only the ones in the damping
functions to be determined independently.

All the developments introduced in this section can be used in a procedure  which can be summarized this way:

\begin{enumerate}
\item Generation of a series of typical condensed-phase configurations
\item DFT calculations on each of these configurations:
\begin{enumerate}
\item Determination of the ground-state wavefunctions, which gives access to the {\it ab initio} forces components
\item Wannier localization, from which the MLWFs spreads and positions are extracted and the {\it ab initio} induced dipoles components are calculated
\end{enumerate}
\item Reconstruction of the electronic density around each ion (following equation \ref{eq:densitywoff})
\item Calculation of the $V^{\rm repulsion}$ term parameters for each ion pair (following equation \ref{eq:repwoff})
\item DFT calculations on each of these configurations under an applied electric field $\mathcal{E}$ (one calculation for each direction), from which new sets of {\it ab initio} induced dipoles components are calculated
\item Calculation of the individual polarizabilities for each ion (following equation \ref{eq:polawoff})
\item Minimization of $\chi_F$ and $\chi_D$  with respect to the damping parameters of the polarization term ($V^{\rm polarization}$) only
\item Determination of the dispersion coefficients for each ion pair (following equations \ref{eq:dispwoff1} and \ref{eq:dispwoff2})
\end{enumerate}

This procedure, which leaves fewer parameters to be fitted than a straightforward application of force-fitting, has up to now been tested and validated in the simple case of NaCl, NaF, KCl and KF.~\cite{rotenberg2010a} The parameters are summarized in table \ref{tab:woffparam} (the polarizabilities are given in table \ref{tab:calcpolarizabilities}). The dynamic (diffusion coefficients, viscosity, thermal and electrical conductivity) and static (density) properties of these molten salts were tested against experimental data: An excellent agreement was obtained, showing again the predictive character of such first-principles based approaches.

\begin{table}[!ht]
\begin{center}
\begin{tabular}{llccccccc}
\hline
System & Ion pair  & $A^{ij}$ & $a^{ij}$ & $C^{ij}_6$ & $C^{ij}_8$ &
$b_D^{ij}$ & $c_D^{ij}$ & $c_D^{ji}$ \\
\hline
KF & F$^-$-F$^-$ & 144.0  & 1.954 & 32.7 & 100.6 & - & - & - \\
   & F$^-$-K$^+$ & 163.3  & 2.052 & 34.3 & 133.1 & 2.052 & 6.4 & -1.2 \\
   & K$^+$-K$^+$ & 109.5  & 2.004 & 36.0 & 168.7 & - & - & - \\
NaF & F$^-$-F$^-$ & 162.9  & 2.012 & 29.4 & 92.1 & - & - & -  \\
    & F$^-$-Na$^+$ &  301.5 & 2.456 & 5.6 & 13.4 & 2.456 & 9.5 & -0.4 \\
    & Na$^+$-Na$^+$ & 599.0  & 3.176 & 1.3 & 2.2 & - & - & - \\
KCl & Cl$^-$-Cl$^-$ & 187.9  & 1.601 & 333.0 & 2516.2 & - & - & - \\
    & Cl$^-$-K$^+$ & 90.2  & 1.636 & 102.4 & 623.6 & 1.636 & 3.8 & 0.4 \\
    & K$^+$-K$^+$ & 244.8  & 2.183 & 35.9 & 166.3 & - & - & - \\
NaCl & Cl$^-$-Cl$^-$ &  205.9 & 1.626 & 298.4 & 2294.1 & - & - & - \\
     & Cl$^-$-Na$^+$ & 94.6  & 1.770 & 14.2 & 66.5 & 1.770 & 3.5 & 0.8 \\
     & Na$^+$-Na$^+$ & 546.4  & 3.1 & 1.3 & 2.3 & - & - & - \\
\hline
\end{tabular}
\end{center}
\caption{ \label{tab:woffparam}
Parameters of the PIM interaction potentials obtained for NaCl, NaF, KCl and KF from the alternative procedure.
For all the ion pairs, we have taken $b_6^{ij}=b_8^{ij}=a^{ij}$. All values are in atomic units.
}
\end{table}

\section{Conclusion}

In conclusion, we have described in this paper two models of different complexity which can be used to describe the interactions in ionic liquids. The first one, the polarizable ion model, includes only one many-body effect, namely the dipole polarization of the ions. This model was recently implemented in the CP2K code.~\cite{cp2k} The second one, the aspherical ion model, not only extends the inclusion of polarization effects up to the quadrupole level, but also includes deformations of the anion in a spherical, dipolar or quadrupolar way for the calculation of the short-range repulsion. Depending on the system of interest and the quantities that have to be determined, one has to choose the most appropriate model.

These interaction potentials include several parameters which have to be determined for each ion type / ion pair. Since the final objective is to predict physico-chemical properties of the materials of interest, no experimental information should be included in the parameterization stage. Here we have summarized the procedure which was used for developing PIM potentials for molten fluorides and AIM potentials for molten oxides and consists in performing an extended force-fitting using {\it ab initio} density functional theory results as reference data. In the last part we show how the use of MLWFs can provide alternative ways for calculating some of the parameters.

Here we have focused on inorganic molten salts. It is worth noticing that an increasing number of studies on room-temperature ionic liquids have involved models similar in spirit to the PIM (they mainly differ due to the choice of Lennard-Jones form to represent the van der Waals interactions).~\cite{yan2004a,chang2009a,yan2010a,bedrov2010a,schroder2010a} For the moment the force-matching procedure was not used in these works, but we have recently extended it to the case of the 1-ethyl-3-methylimidazolium tetrachloroaluminate  ionic liquid.~\cite{salanneinpress}

\section*{Acknowledgements}
S.J. was supported by the German Science Foundation (DFG) under grant 
no. JA 1469/4-1. M.S., B.R., R.V. and C.S. acknowledge the support of the French Agence Nationale de la
Recherche (ANR), under grant SYSCOMM (ANR-09-SYSC-012) "Multiscale
Simulation of Ions at Solid-Liquid Interfaces", and of PACEN (Programme sur l'Aval du Cycle Nucl\'eaire) through PCR ANSF and GNR PARIS programs. We would like to thank Sami Tazi, Toon Verstraelen and Teodoro Laino for their help in implementing the PIM in the CP2K code.

\end{document}